\documentclass[final, 3p, twocolumn, times, 10pt]{elsarticle}

\usepackage{amssymb}
\biboptions{sort}

\usepackage[dvipsnames]{xcolor}
\usepackage{fancyhdr}
\usepackage[utf8]{inputenc}
\usepackage[us,12hr]{datetime}
\usepackage[hyphens]{url}
\usepackage{microtype}
\usepackage[keeplastbox, ancient]{flushend}
\usepackage[binary-units]{siunitx}
\usepackage{setspace}

\journal{Microprocessors and Microsystems}

\begin{document}

\makeatletter
\def\ps@pprintTitle{%
   \let\@oddhead\@empty
   \let\@evenhead\@empty
   \def\@oddfoot{\reset@font\hfil\thepage\hfil}
   \let\@evenfoot\@oddfoot
}
\makeatother

\begin{frontmatter}

\title{Processing Data Where It Makes Sense:\\Enabling In-Memory Computation}

\author[1,2]{Onur Mutlu}
\author[2]{Saugata Ghose}
\author[1]{Juan G\'omez-Luna}
\author[2,3]{Rachata Ausavarungnirun}

\address[1]{ETH Z\"urich}
\address[2]{Carnegie Mellon University}
\address[3]{King Mongkut's University of Technology North Bangkok\vspace{-10pt}}

\begin{abstract}
Today's systems are overwhelmingly designed to move data to
computation. This design choice goes directly against at least three
key trends in systems that cause performance, scalability and energy
bottlenecks: (1)~data access from memory is already a key bottleneck
as applications become more data-intensive and memory bandwidth and
energy do not scale well, (2)~energy consumption is a key constraint
in especially mobile and server systems, (3)~data movement is very
expensive in terms of bandwidth, energy and latency, much more so than
computation. These trends are especially severely-felt in the
data-intensive server and energy-constrained mobile systems of today.

At the same time, conventional memory technology is facing many
scaling challenges in terms of reliability, energy, and
performance. As a result, memory system architects are open to
organizing memory in different ways and making it more intelligent, at
the expense of higher cost. The emergence of 3D-stacked memory plus
logic as well as the adoption of error correcting codes inside DRAM
chips, {and the necessity for designing new solutions to serious
  reliability and security issues, such as the RowHammer phenomenon,}
are an evidence of this trend.
 
In this work, we discuss some recent research that aims to practically
enable computation close to data. After motivating trends in
applications as well as technology, we discuss at least two promising
directions for \emph{processing-in-memory} (PIM): (1)~performing
massively-parallel bulk operations in memory by exploiting the analog
operational properties of DRAM, with low-cost changes, (2)~exploiting
the logic layer in 3D-stacked memory technology to accelerate
important data-intensive applications. In both approaches, we describe
and tackle relevant cross-layer research, design, and adoption
challenges in devices, architecture, systems, and programming
models. Our focus is {on} the development of in-memory processing
designs that can be adopted in real computing platforms at low cost.
\end{abstract}

\begin{keyword}
data movement \sep main memory \sep processing-in-memory \sep 3D-stacked memory \sep {near-data processing}
\end{keyword}

\end{frontmatter}

\section{Introduction}
\label{sec:introduction}

Main memory, which is built using the Dynamic Random Access Memory
(DRAM) technology, is a major component in nearly all computing
systems. Across all of these systems, including servers, cloud
platforms, and mobile/embedded devices, the {data} working set
sizes of modern applications are rapidly growing, causing the main
memory to be a significant bottleneck for these
applications~\cite{mutlu.imw13, mutlu.superfri15,dean.cacm13,
  kanev.isca15, ferdman.asplos12, wang.hpca14, boroumand.asplos18}.
Alleviating the main memory bottleneck requires the memory capacity,
energy, cost, and performance to all scale in an efficient manner.
Unfortunately, it has become increasingly difficult in recent years to
scale all of these dimensions{~\cite{kang.memoryforum14, mutlu.imw13,
  mutlu.superfri15, mckee.cf04,
  wilkes.can01,kim-isca2014,salp,yoongu-thesis,raidr,mutlu2017rowhammer,donghyuk-ddma,lee-isca2009,rbla,yoon-taco2014,lim-isca09,
  wulf1995hitting, chang.sigmetrics16, lee.hpca13, lee.hpca15,
  chang.sigmetrics17, lee.sigmetrics17, luo.dsn14,
  luo.arxiv17,hassan2017softmc,chargecache, kevinchang-thesis}}, and the main memory
bottleneck has instead been worsening.

A major reason for the main memory bottleneck is the high cost
associated with \emph{data movement}.  In today's computers, to
perform any operation on data that resides in main memory, the memory
controller must first issue a series of commands to the DRAM modules
across an off-chip bus (known as the \emph{memory channel}).  The DRAM
module responds by sending the data to the memory controller across
the memory channel, after which the data is placed within a cache
{or registers}.  The CPU can only perform the operation on the
data once the data is in the cache.  This process of moving data from
the DRAM to the CPU incurs a long latency, and consumes a significant
amount of energy~\cite{hashemi.isca16,cont-runahead,
  ahn.tesseract.isca15, ahn.pei.isca15,boroumand.asplos18}.  These
costs are often exacerbated by the fact that much of the data brought
into the caches is \emph{not reused} by the
CPU~\cite{qureshi.isca07,qureshi-hpca07}, providing little benefit in
return for the high latency and energy cost.

The cost of data movement is a fundamental issue with the
\emph{processor-centric} nature of contemporary computer systems,
where the CPU is considered to be the master of the system and
{has been optimized heavily}.  In contrast, data storage units
such as main memory are treated as unintelligent workers, and, thus,
are largely not optimized.  With the increasingly \emph{data-centric}
nature of contemporary and emerging applications, the
processor-centric design approach leads to many inefficiencies.  For
example, {within a single compute node, most of the node real
  estate is dedicated} to handle the storage and movement of data
(e.g., large on-chip caches, shared interconnect, memory
controllers, off-chip {interconnects}, main memory)~\cite{kumar.isscc2009}.

Recent advances in memory design {and memory architecture} have
enabled the opportunity for a paradigm shift towards performing
\emph{processing-in-memory} (PIM), where we can redesign the computer
to no longer be processor-centric and avoid unnecessary data movement.
Processing-in-memory, also known as \emph{near-data processing} (NDP),
enables the ability to perform operations either using (1)~the memory
itself, or (2)~some form of processing logic (e.g., accelerators,
simple cores, reconfigurable logic) inside the DRAM subsystem.
Processing-in-memory has been proposed for {at least} four
decades{~\cite{stone1970logic,shaw1981non,
    elliott1992computational, kogge1994execube, gokhale1995processing,
    patterson1997case, oskin1998active, kang2012flexram,
    Draper:2002:ADP:514191.514197, Mai:2000:SMM:339647.339673,
    elliott.dt99, riedel.1998, keeton.1998, kaxiras.1997,
    acharya.1998}}.
    {However, these past efforts were}
{\emph{not}} adopted at large scale due to {various reasons,
  including} the difficulty of integrating processing elements with
DRAM {and the fact that memory technology was not facing as
  critical scaling challenges as it is today}.  As a result of
advances in modern memory architectures, {e.g., the integration
  of logic and memory in a 3D-stacked manner}, various recent works
explore a range of PIM architectures for multiple different purposes
(e.g., \cite{zhu2013accelerating, pugsley2014ndc, zhang.hpdc14,
  farmahini-farahani.hpca15, ahn.tesseract.isca15, ahn.pei.isca15,
  loh2013processing, hsieh.isca16, pattnaik.pact16,
  DBLP:conf/isca/AkinFH15, impica, DBLP:conf/sigmod/BabarinsaI15,
  DBLP:conf/IEEEpact/LeeSK15, DBLP:conf/hpca/GaoK16, chi.isca16,
  gu.isca16, kim.isca16, asghari-moghaddam.micro16, boroumand2016pim,
  hashemi.isca16, cont-runahead, GS-DRAM, liu-spaa17, gao.pact15,
  guo2014wondp, sura.cf15, morad.taco15, hassan.memsys15, li.dac16,
  kang.icassp14, aga.hpca17, shafiee.isca16, seshadri2013rowclone,
  Seshadri:2015:ANDOR, chang.hpca16, seshadri.arxiv16,
  seshadri.micro17, nai2017graphpim,kim.arxiv17,kim.bmc18, li.micro17,
  kim.sc17, boroumand.asplos18}).

In this paper, we explore two approaches to enabling
processing-in-memory in modern systems.  The first approach examines a
form of PIM that only \emph{minimally changes memory chips} to perform
simple yet powerful common operations that the chip could be made
inherently very good at performing~\cite{seshadri.bookchapter17,
  seshadri2013rowclone,
  chang.hpca16,kevinchang-thesis,seshadri.thesis16,
  Seshadri:2015:ANDOR, seshadri.arxiv16, seshadri.micro17, li.micro17,
  GS-DRAM, ghose.bookchapter19, ghose.bookchapter19.arxiv,
  deng.dac2018}.  Solutions that fall under
this approach take advantage of the existing DRAM design to cleverly
and efficiently perform \emph{bulk operations} (i.e., operations on an
entire row of DRAM cells), such as bulk copy, data initialization, and
bitwise operations.  The second approach takes advantage of the design
of emerging \emph{3D-stacked memory technologies} to enable PIM in
{a more} general-purpose manner~\cite{loh2013processing, pugsley2014ndc,
  zhu2013accelerating, DBLP:conf/isca/AkinFH15,impica,
  ahn.tesseract.isca15, nai2017graphpim,
  DBLP:conf/sigmod/BabarinsaI15, gao.pact15, kim.bmc18, kim.arxiv17,
  gu.isca16, boroumand2016pim, boroumand.asplos18, chi.isca16,
  kim.isca16, DBLP:conf/IEEEpact/LeeSK15, ahn.pei.isca15,
  zhang.hpdc14, hsieh.isca16, pattnaik.pact16, sura.cf15,
  hassan.memsys15,farmahini-farahani.hpca15, DBLP:conf/hpca/GaoK16,
  guo2014wondp,liu-spaa17,kim.sc17}.  In order to stack multiple
layers of memory, 3D-stacked chips use vertical \emph{through-silicon
  vias} (TSVs) to connect the layers to each other, and to the I/O
drivers of the chip~\cite{lee.taco16}.  The TSVs provide much greater
\emph{internal} bandwidth than is available externally on the memory
channel.  Several such 3D-stacked memory architectures, such as the
Hybrid Memory Cube~\cite{hmc.spec.1.1, hmc.spec.2.0} and
High-Bandwidth Memory~\cite{jedec.hbm.spec,lee.taco16}, include a
\emph{logic layer}, where designers can add some simple processing
logic to take advantage of the high internal bandwidth.

For both approaches to PIM, there are a number of new challenges that
system architects and programmers must address to enable the
widespread adoption of PIM across the computing landscape and in
different domains of workloads.  In addition to describing work along
the two key approaches, we also discuss these challenges in this
paper, along with existing work that addresses these challenges.

\section{Major Trends Affecting Main Memory}
\label{sec:majortrends}

The main memory is a major, critical component of all computing
systems, including cloud and server platforms, desktop computers,
mobile and embedded devices, and sensors. It is one of the two main
pillars of any computing platform, together with the processing
elements, namely CPU cores, GPU cores, or reconfigurable devices.

Due to its relatively low cost and low latency, DRAM is the
predominant technology to build main memory. Because of the growing
data working set sizes of modern applications~\cite{mutlu.imw13,
  mutlu.superfri15,dean.cacm13, kanev.isca15, ferdman.asplos12,
  wang.hpca14, boroumand.asplos18}, there is an ever-increasing demand
for higher DRAM capacity and performance. Unfortunately, DRAM
technology scaling is becoming more and more challenging in terms of
increasing the DRAM capacity and maintaining the DRAM energy
efficiency and
reliability~\cite{liu.isca13,mutlu.imw13,kim-isca2014,mutlu2017rowhammer,ghose.sigmetrics18}. Thus,
fulfilling the increasing memory needs from applications is becoming
more and more costly {and
  difficult}{~\cite{dean.cacm13,kanev.isca15,mckee.cf04,mutlu.superfri15,wilkes.can01,salp,kang.memoryforum14,yoongu-thesis,raidr,ahn.tesseract.isca15,ahn.pei.isca15,hsieh.isca16,donghyuk-ddma,lee-isca2009,rbla,yoon-taco2014,lim-isca09,
  wulf1995hitting,chang.sigmetrics16,lee.hpca13,lee.hpca15,chang.sigmetrics17,lee.sigmetrics17,luo.dsn14,luo.arxiv17,hassan2017softmc,chargecache,wang.micro18,das.dac18,kim.iccd18, kevinchang-thesis}}.

If CMOS technology scaling is coming to an end~\cite{denning.2016},
the projections are significantly worse for DRAM technology
scaling~\cite{itrs.2009}. DRAM technology scaling affects {all
  major} characteristics of DRAM, {including} capacity,
bandwidth, latency, {energy and cost}.  We next describe the key
{issues and trends in DRAM technology scaling} and discuss how
these trends motivate the need for intelligent memory controllers,
which can be used as a substrate for processing in memory.

The first key {concern is the difficulty} of scaling DRAM
capacity (i.e., density, or cost per bit), bandwidth and latency {\em
  at the same time}.  While the processing core count doubles every
two years, the DRAM capacity doubles only every three
years~\cite{lim-isca09}. This causes the \emph{memory capacity per
  core} to drop by approximately 30\% every two
years~\cite{lim-isca09}. The trend is even worse for \emph{memory
  bandwidth per core} -- in the last 20 years, {DRAM chip
  capacity (for the most common DDRx chip of the time)} has improved
around 128$\times$ while DRAM bandwidth has increased only around
20$\times$~\cite{kevinchang-thesis,chang.sigmetrics16,lee.hpca13}. In
the same period of twenty years, DRAM latency (as measured by the row
cycling time) has remained almost constant (i.e., reduced by only
30\%), making it a significant performance bottleneck for {many
  modern workloads, including} in-memory
databases~\cite{ailamaki1999dbmss,boncz.1999,clapp.2015,JAFAR}, graph
processing~\cite{ahn.tesseract.isca15,umuroglu.2015,xu.2014}, data
analytics~\cite{awan.2015,awan.2016,clapp.2015,yasin.2014}, datacenter
workloads~\cite{kanev.isca15}, {and consumer
  workloads~\cite{boroumand.asplos18}}. As low-latency computing is
becoming ever more important~\cite{mutlu.imw13}, e.g., due to the
ever-increasing need to process large amounts of data at real time,
and predictable performance continues to be a critical concern in the
design of modern computing
systems~\cite{moscibroda-usenix2007,mutlu-micro2007,mutlu-isca2008,mutlu.superfri15,lavanya-thesis,
  donghyuk-ddma,subramanian-hpca2013,usui-taco2016,subramanian-micro2015},
it is increasingly critical to design low-latency main memory chips.

The second key {concern is that DRAM technology scaling to
  smaller nodes adversely} affects DRAM reliability.  A DRAM cell
stores {each bit} in the form of charge in a capacitor, which is
accessed via an access transistor and peripheral circuitry. For a DRAM
cell to operate correctly, both the capacitor and the access
transistor (as well as the peripheral circuitry) need to operate
reliably. As the size of the DRAM cell reduces, both the capacitor and
the access transistor become less reliable. As a result,
{reducing the size of the DRAM cell} increases the difficulty of
correctly storing and detecting the desired original value in the DRAM
cell~\cite{liu.isca13,mutlu.imw13,kim-isca2014,mutlu2017rowhammer}. Hence,
memory scaling causes memory errors to appear more frequently. For
example, a study of Facebook's entire production datacenter servers
showed that memory errors, and thus the server failure rate, increase
proportionally with the density of the chips employed in the
servers~\cite{meza.dsn15}. Thus, it is critical to make the main memory
system more reliable to build reliable computing systems on top of it.

The third key {issue} is that the reliability problems caused by
aggressive DRAM technology scaling can leads to new security
vulnerabilities. {The RowHammer
  phenomenon}~\cite{kim-isca2014,mutlu2017rowhammer} shows that it is
possible to {predictably} induce errors {(bit flips)} in
most {modern} DRAM chips.  {Repeatedly reading the same row
  in DRAM can corrupt data in physically-adjacent rows. Specifically,
  when a DRAM row is opened (i.e., activated) and closed (i.e.,
  precharged) repeatedly (i.e., hammered), enough times within a DRAM
  refresh interval, one or more bits in physically-adjacent DRAM rows
  can be flipped to the wrong value.  A very simple user-level
  program~\cite{rowhammer.github} can reliably and consistently induce
  RowHammer errors in vulnerable DRAM modules.}  {The seminal
  paper that introduced RowHammer~\cite{kim-isca2014} showed that more
  than 85\% of the chips tested, built by three major vendors between
  2010 and 2014, were vulnerable to RowHammer-induced errors}. In
particular, \emph{all} DRAM modules from 2012 and 2013 are vulnerable.

The RowHammer {phenomenon} entails a real reliability, and
perhaps even more importantly, a real and prevalent security issue. It
breaks physical {memory} isolation between two addresses, one of
the fundamental building blocks of memory, on top of which system
security principles are built. With RowHammer, accesses to one row
(e.g., an application page) can modify data stored in another memory
row (e.g., an OS page).  This was confirmed by researchers from Google
Project Zero, who developed a user-level attack that uses RowHammer to
gain kernel privileges~\cite{seaborn.2015,seaborn.2016}.  Other
researchers have shown how RowHammer vulnerabilities can be exploited
in various ways to gain privileged access to various systems: in a
remote server RowHammer can be used to remotely take over the server
via the use of JavaScript~\cite{gruss.2015}; a virtual machine can
take over another virtual machine by inducing errors in the victim
virtual machine's memory space~\cite{razavi.2016}; a malicious
application without permissions can take control of an Android mobile
device~\cite{vanderveen.2016}; or an attacker can gain arbitrary
read/write access in a web browser on a Microsoft Windows 10
system~\cite{bosman.2016}. For a more detailed treatment of the
RowHammer problem and its consequences, we refer the reader
to~\cite{kim-isca2014,mutlu2017rowhammer,rowhammer-topinhes18}.

The fourth key {issue} is the power and energy consumption of
main memory. DRAM is inherently a power and energy hog, as it consumes
energy even when it is not used (e.g., it requires periodic memory
refresh~\cite{raidr}), due to its charge-based nature. And, energy
consumption of main memory is becoming worse due to three major
reasons. First, its capacity and complexity are both
increasing. Second, main memory has remained off the main processing
chip, even though many other platform components have been integrated
into the processing chip and have benefited from the aggressive energy
scaling and low-energy communication substrate on-chip. Third, the
difficulties in DRAM technology scaling are making energy reduction
very difficult with technology generations. For example,
{Lefurgy et al.~\cite{lefurgy.2003} showed, in 2003 that, in
  large commercial servers designed by IBM, the off-chip memory
  hierarchy (including, at that time, DRAM, interconnects, memory
  controller, and off-chip caches) consumed between 40\% and 50\% of
  the total system energy. The trend has become even worse over the
  course of the one-to-two decades.}  In recent computing systems with
CPUs or GPUs, {only DRAM itself is shown to account for} more
than 40\% of the total system power~\cite{ware.2010,paul.2015}. Hence,
the power and energy consumption of main memory is increasing relative
to that of other components in computing platform. As energy
efficiency and sustainability are critical necessities in computing
platforms today, it is critical to reduce the energy and power
consumption of main memory.

\section{The Need for Intelligent Memory Controllers to Enhance Memory Scaling}
\label{sec:intmemcont}

A key promising approach to solving the four major issues above is to
design {\em intelligent memory controllers} that can manage main
memory better. If the memory controller is designed to be more
intelligent and more programmable, it can, for example, incorporate
flexible mechanisms to overcome various types of reliability issues
(including RowHammer), manage latencies and power consumption better
based on a deep understanding of the DRAM and application
characteristics, provide enough support for programmability to prevent
security and reliability vulnerabilities that are discovered in the
field, and manage various types of memory technologies that are put
together as a hybrid main memory to enhance the scaling of the main
memory system. We provide a few examples of how an intelligent memory
controller can help overcome circuit- and device-level issues we are
facing at the main memory level. We believe having intelligent memory
controllers can greatly alleviate the scaling issues encountered with
main memory today, as we have described in an earlier position
paper~\cite{mutlu.imw13}. This is a direction that is also supported
by industry today, as described in an informative paper written
collaboratively by Intel and Samsung engineers on DRAM technology
scaling issues{~\cite{kang.memoryforum14}}.

First, the RowHammer vulnerability can be prevented by
probabilistically refreshing rows that are adjacent to an activated
row, with a very low probability. This solution, called PARA
(Probabilistic Adjacent Row Activation)~\cite{kim-isca2014} was shown
to provide strong, programmable guarantees against RowHammer, with
very little power, performance and chip area
overhead~\cite{kim-isca2014}. It requires a slightly more intelligent
memory controller that knows (or that can figure out) the physical
adjacency of rows in a DRAM chip and that is programmable enough to
adjust the probability of adjacent row activation and issue refresh
requests to adjacent rows accordingly to the probability supplied by
the system. As described by prior
work~\cite{kim-isca2014,mutlu2017rowhammer,rowhammer-topinhes18}, this
solution is much lower overhead that {increasing the} refresh rate across
the board for the entire main memory, which is the RowHammer solution
employed by existing systems in the field that have simple and rigid
memory controllers.

Second, an intelligent memory controller can greatly alleviate the
refresh problem in DRAM, and hence its negative consequences on
energy, performance, predictability, and technology scaling, by
understanding the retention time characteristics of different rows
well. It is well known that the retention time of different cells in
DRAM are widely different due to process manufacturing
variation~\cite{raidr,liu.isca13}. Some cells are strong (i.e., they
can retain data for hundreds of seconds), whereas some cells are weak
(i.e., they can retain data for only \SI{64}{\milli\second}). Yet, today's memory
controllers treat {every cell} as equal and refresh all rows every \SI{64}{\milli\second},
which is the {worst}-case retention time that is allowed. This
worst-case refresh rate leads to a large number of unnecessary
refreshes, and thus great energy waste and performance loss. Refresh
is also shown to be the key technology scaling limiter of
DRAM~\cite{kang.memoryforum14}, and as such refreshing all DRAM cells
at the worst case rates is likely to make DRAM technology scaling
difficult.  An intelligent memory controller can overcome the refresh
problem by identifying the minimum data retention time of each row
(during online operation) and refreshing each row at the rate it
really requires to be refreshed at or by {decommissioning} weak rows
such that data is not stored in them. As shown by a recent body of
work whose aim is to design such an intelligent memory controller that
can perform inline profiling of DRAM cell retention times and online
adjustment of refresh rate on a per-row
basis~\cite{raidr,liu.isca13,khan.sigmetrics14,qureshi.dsn15,khan.dsn16,khan.cal16,patel.isca17,khan.micro17},
including the works on RAIDR~\cite{raidr,liu.isca13},
AVATAR~\cite{qureshi.dsn15} and REAPER~\cite{patel.isca17}, such an
intelligent memory controller can eliminate more than 75\% of all
refreshes at very low cost, leading to significant energy reduction,
performance improvement, and quality of service benefits, all at the
same time. Thus the downsides of DRAM refresh can potentially be
overcome with the design of intelligent memory controllers.

{Third, an intelligent memory controller can enable performance
  improvements that can overcome the limitations of memory scaling.
  As we discuss in Section~\ref{sec:majortrends}, DRAM latency has
  remained almost constant over the last twenty years, despite the
  fact that low-latency computing has become more important during
  that time. Similar to how intelligent memory controllers handle the
  refresh problem, the controllers can exploit the fact that not all
  cells in DRAM need the same amount of time to be accessed.
  Manufacturers assign timing parameters that define the amount of
  time required to perform a memory access.  In order to guarantee
  correct operation, the timing parameters are chosen to ensure that
  the \emph{worst-case} cell in any DRAM chip that is sold can still
  be accessed correctly at \emph{worst-case operating
    temperatures}~\cite{chang.sigmetrics16, lee.hpca15,
    lee.sigmetrics17, kim.iccd18}. However, we find that access
  latency to cells is very heterogeneous due to variation in both
  operating conditions (e.g., across different temperatures and
  operating voltage), manufacturing process (e.g., across different
  chips and different parts of a chip), and access patterns (e.g.,
  whether or not the cell was recently accessed). We give six examples
  of how an intelligent memory controller can exploit the various
  different types of heterogeneity.

  {(1)~At} low temperature, DRAM cells contain more charge, and
  as a result, can be accessed much faster than at high
  temperatures. We find that, averaged across 115 real DRAM modules
  from three major manufacturers, read and write latencies of DRAM can
  be reduced by 33\% and {55\%, respectively,} when operating at
  relatively low temperature ({\SI{55}{\celsius}}) compared to
  operating at worst-case temperature
  ({\SI{85}{\celsius}}){~\cite{lee.hpca15, lee.thesis16}}. Thus, a slightly
  intelligent memory controller can greatly reduce memory latency by
  adapting the access latency to operating temperature.

  {(2)~Due} to manufacturing process variation, we find that the
  majority of cells in DRAM (across different chips or within the same
  chip) can be accessed much faster than the manufacturer-provided
  timing parameters{~\cite{chang.sigmetrics16, lee.hpca15,
    lee.sigmetrics17, kim.iccd18, kevinchang-thesis, lee.thesis16}}.  
  An intelligent memory controller
  can profile the DRAM chip and identify which cells can be accessed
  reliably at low latency, and use this information to reduce access
  latencies by as much as 57\%~\cite{chang.sigmetrics16,
    lee.sigmetrics17, kim.iccd18}.

  {(3)~In} a similar fashion, an intelligent memory controller
  can use similar properties of manufacturing process variation to
  reduce the energy consumption of a computer system, by exploiting
  the minimum voltage required for safe operation of different parts
  of a DRAM chip{~\cite{chang.sigmetrics17, kevinchang-thesis}}. 
  The key idea is to reduce
  the operating voltage of a DRAM chip from the standard specification
  and tolerate the resulting errors by increasing access latency on a
  per-bank basis, while keeping performance degradation in check.

  {(4)~Bank} conflict latencies can be dramatically reduced by
  making modifications in the DRAM chip such that different subarrays
  in a bank can be accessed mostly independently{,} and designing
  an intelligent memory controller that can take {advantage} of
  requests that require data from different subarrays (i.e., exploit
  subarray-level parallelism){~\cite{salp, yoongu-thesis}.}

  {(5)~Access} latency to a portion of the DRAM bank can be
  greatly reduced by partitioning the DRAM array such that a subset of
  rows can be accessed much faster than the other rows and having an
  intelligent memory controller that decides what data should be
  placed in fast rows versus slow rows{~\cite{lee.hpca13, lee.thesis16}}.

  {(6)~We} find that a {recently-accessed or
    recently-refreshed} memory row can be accessed much more quickly
  than the standard latency if it needs to be accessed again soon,
  since the recent access and refresh to the row has replenished the
  charge of the cells in the row. An intelligent memory controller can
  thus keep track of the charge level of recently-accessed/refreshed
  rows and use the appropriate access latency that corresponds to the
  charge level~\cite{chargecache,wang.micro18,das.dac18}, leading to
  significant reductions in both access and refresh latencies.
  {Thus,} the poor scaling of DRAM latency and energy can
  potentially be overcome with the design of intelligent memory
  controllers that can facilitate a large number of effective latency
  and energy reduction techniques.}

{Intelligent controllers are already in widespread use in another
  key part of a modern computing system.  In solid-state drives (SSDs)
  consisting of NAND flash memory, the flash controllers that manage
  the SSDs are designed to incorporate a significant level of
  intelligence in order to improve both performance and
  reliability{~\cite{cai.procieee17, yucai.bookchapter18, cai.bookchapter18.arxiv,
    tavakkol.fast18, tavakkol.isca18}}.  Modern flash controllers need
  to take into account a wide variety of issues such as remapping
  data, performing wear leveling to mitigate the limited lifetime of
  NAND flash memory devices, refreshing data based on the current
  wearout of each NAND flash cell, optimizing voltage levels to
  maximize memory lifetime, and enforcing fairness across different
  applications accessing the SSD.  Much of the complexity in flash
  controllers is a result of mitigating issues related to the scaling
  of NAND flash memory{~\cite{cai.procieee17, yucai.bookchapter18, 
  cai.bookchapter18.arxiv, yucai-thesis, luo.thesis18}}.  
  We argue that in order to overcome scaling
  issues in DRAM, the time has come for DRAM memory controllers to
  also incorporate significant intelligence.}

{As we describe above, introducing intelligence into the memory
  controller can help us overcome a number of key challenges in memory
  scaling.  In particular, a significant body of works have
  demonstrated that the key reliability, refresh{,} and latency/energy
  issues in memory can be mitigated effectively with an intelligent
  memory controller.  As we discuss {in Section~\ref{sec:processorcentric}}, this
  intelligence can go even further, by enabling the memory controllers
  (and the broader memory system) to perform application computation
  in order to overcome the significant data movement bottleneck in
  existing computing systems.}

\section{Perils of Processor-Centric Design}
\label{sec:processorcentric}

A major bottleneck against improving the overall system performance
and the energy efficiency of today's computing systems is the high
cost of \emph{data movement}. This is a natural consequence of the von
Neumann model~\cite{burks.1946}, which separates computation and
storage in two different system components (i.e., the computing
{unit versus the memory/storage unit}) that are connected by an
off-chip bus. With this model, processing is done only in one place,
while data {is} stored in another, {separate} place. Thus,
data needs to move back and forth between the memory/storage unit and
{the computing unit (e.g., CPU cores or accelerators)}.

In order to perform an operation on data that is stored within memory,
a costly process is invoked. First, the CPU {(or an
  accelerator)} must issue a request to the memory controller, which
in turn sends a series of commands across the off-chip bus to the DRAM
module. Second, the data is read from the DRAM module and returned to
the memory controller. Third, the data is placed in the CPU cache
{and registers}, where it is accessible by the CPU
cores. Finally, the CPU can operate (i.e., perform computation) on the
data. All these steps consume substantial time and energy in order to
bring the data into the CPU
chip~\cite{kestor.iiswc2013,pandiyan.iiswc2014,kanev.isca15,boroumand.asplos18}.

{In current computing systems, the CPU is the only system
  component that is able to perform computation on data. The rest of
  system components are devoted to only data storage (memory, caches,
  disks) and data movement (interconnects); they are incapable of
  performing computation.} As a result, current computing systems are
\emph{grossly imbalanced}, {leading to} large amounts of energy
inefficiency and low performance. As empirical evidence to the gross
imbalance caused by the processor-memory dichotomy in the design of
computing systems today, we have recently observed that more than 62\%
of the entire system energy consumed by four major commonly-used
mobile consumer workloads (including the Chrome browser, TensorFlow
machine learning inference engine, and the VP9 video encoder and
decoder)~\cite{boroumand.asplos18}. Thus, the fact that current
systems can perform computation only in the computing unit (CPU cores
and hardware accelerators) is causing significant waste in terms of
energy by necessitating data movement across the entire system.

At least five factors contribute to the performance loss and the
energy waste associated with retrieving data from main memory, which
we briefly describe next.

First, the width of the off-chip bus between the memory controller and
the main memory is narrow, due to pin count and {cost} constraints,
leading to relatively low bandwidth to/from main memory. This makes
{it} difficult to send a large number of requests to memory in
parallel.

Second, current computing systems deploy complex multi-level cache
hierarchies and latency tolerance/hiding mechanisms (e.g.,
{sophisticated caching algorithms at many different caching
  levels,} multiple complex prefetching techniques, {high amounts
  of multithreading, complex out-of-order execution}) to tolerate the
data access from memory. These components, while sometimes effective
at improving performance, are costly in terms of both die area and
energy consumption, {as well as the additional latency required
  to access/manage them}. These components also increase the
complexity of the system significantly. Hence, the architectural
techniques used in modern systems to tolerate the consequences of the
dichotomy between processing unit and main memory, lead to significant
energy waste and additional complexity.

Third, the caches are not always properly leveraged. Much of the data
brought into the caches is \emph{not} reused by the
CPU~\cite{qureshi.isca07,qureshi-hpca07}, e.g., {in} streaming
or random access applications. This renders the caches either very
inefficient or unnecessary {for a wide variety of modern
  workloads}.

Fourth, many modern applications, such as graph
processing~\cite{ahn.tesseract.isca15,ahn.pei.isca15}, produce random
memory access patterns. In such cases, not only the caches but also
the off-chip bus and the DRAM memory itself {become} very
inefficient, since only a little part of each cache line retrieved is
actually used by the CPU. Such accesses are also not easy to prefetch
and often either confuse the prefetchers or render them ineffective.
{Modern memory hierarchies are not designed to work well for
  random access {patterns}.}

Fifth, the computing unit and the memory unit are connected through
long, power-hungry interconnects. {These interconnects impose
  significant additional latency to every data access and represent a
  significant fraction of the energy spent on moving data to/from the
  DRAM memory.} In fact, off-chip interconnect latency and energy
consumption is a key limiter of performance and energy in modern
systems~\cite{lee.hpca13,donghyuk-ddma,seshadri2013rowclone,GS-DRAM}
as it greatly exacerbates the cost of data movement.

The increasing disparity between processing technology and
memory/communication technology has resulted in systems in which
communication (data movement) costs dominate computation costs in
terms of energy consumption. The energy consumption of a main memory
access is between two to three orders of magnitude the energy
consumption of a complex addition operation today.  For
example, \cite{pandiyan.iiswc2014} reports that the energy consumption
of a memory access is $\sim115\times$ the energy consumption of an
addition operation. As a result, data movement accounts for
40\%~\cite{kestor.iiswc2013}, 35\%~\cite{pandiyan.iiswc2014}, and
62\%~\cite{boroumand.asplos18} of the total system energy in
scientific, mobile, and consumer applications, respectively.  This
energy waste due to data movement is a huge burden that greatly limits
the efficiency and performance of all modern computing platforms, from
datacenters with a restricted power budget to mobile devices with
limited {battery life}.

Overcoming all the reasons that cause low performance and large energy
inefficiency (as well as high system design complexity) in current
computing systems requires a paradigm shift. {We believe that}
future computing architectures should become more \emph{data-centric}:
{they should (1) perform computation with} minimal data
movement, and (2) {compute} where it makes sense (i.e., where
the data resides), as opposed to computing solely in the CPU or
accelerators. Thus, the traditional rigid dichotomy between the
computing units and the memory/communication units needs to be broken
and a new paradigm enabling computation where the data resides needs
to be invented and enabled.

\section{Processing-in-Memory (PIM)}
\label{sec:pim}

{Large amounts of data movement is a major result of the
  predominant processor-centric design paradigm of modern
  computers. Eliminating unnecessary data movement between memory unit
  and compute unit is essential to make future computing architectures
  higher performance, more energy efficient and sustainable. To this
  end, \emph{processing-in-memory} (PIM) equips the memory {subsystem}
  with the ability to perform computation.}

In this section, we describe two promising approaches to implementing
PIM in modern architectures.  The first approach exploits the existing
DRAM architecture and the operational principles of the DRAM circuitry
to enable bulk processing operations within the main memory with
minimal changes. This minimalist approach can especially be powerful
in performing specialized computation in main memory by taking
advantage of what the main memory substrate is extremely good at
performing with minimal changes to the existing memory chips.  The
second approach exploits the ability to implement a wide variety of
general-purpose processing logic in the logic layer of 3D-stacked
memory and thus the high internal bandwidth {and low latency}
available {between the logic layer and the memory layers of}
3D-stacked memory. This is a more general approach where the logic
implemented in the logic layer can be general purpose and thus can
benefit a wide variety of applications.

\subsection{Approach I: Minimally Changing Memory Chips}
\label{sec:minimally}

One approach to {implementing} processing-in-memory
{modifies existing DRAM architectures minimally to extend their
  functionality with computing capability.  This approach takes
  advantage of the existing interconnects in and analog operational
  behavior of conventional DRAM architectures (e.g., DDRx, LPDDRx,
  HBM), without the need for a dedicated logic layer or logic
  processing elements}, and {usually} with very low overheads.
Mechanisms that use {this approach} take advantage of the high
internal bandwidth available within {each} DRAM cell
array. There are a number of example PIM architectures that make use
of this
approach~\cite{seshadri.bookchapter17,seshadri2013rowclone,chang.hpca16,kevinchang-thesis,seshadri.thesis16,
  Seshadri:2015:ANDOR, seshadri.arxiv16, seshadri.micro17}.  In this
section, we first focus on two such {designs}: {RowClone,
  which enables in-DRAM bulk data movement
  operations~\cite{seshadri2013rowclone} and Ambit, which enables
  in-DRAM} bulk bitwise operations~\cite{Seshadri:2015:ANDOR,
  seshadri.arxiv16, seshadri.micro17}. Then, we describe a low-cost
substrate that {performs} data reorganization for non-unit
strided access patterns~\cite{GS-DRAM}.

\subsubsection{RowClone}
\label{sec:rowclone}

Two important classes of bandwidth-intensive memory operations are
(1)~\emph{bulk data copy}, where a large quantity of data is copied
from one location in physical memory to another; and (2)~\emph{bulk
  data initialization}, where a large quantity of data is initialized
to a specific value. We refer to these two operations as \emph{bulk
  data movement operations}. Prior
research{~\cite{os-hardware,arch-os, kanev.isca15}} has shown that
operating systems {and data center workloads} spend a significant
portion of their time performing bulk data movement
operations. Therefore, accelerating these operations will likely
improve system performance {and energy} efficiency.

We have developed a mechanism called
\emph{RowClone}~\cite{seshadri2013rowclone}, which takes advantage of
the fact that bulk data movement operations do \emph{not} require any
computation on the part of the processor.  RowClone exploits the
internal organization and operation of DRAM to perform bulk data
copy/initialization quickly and efficiently inside {a DRAM chip}.
A DRAM chip contains multiple banks, where the banks are connected
together and to I/O circuitry by a shared internal bus, each of which
is divided into multiple
\emph{subarrays}~\cite{salp,chang.hpca14,seshadri2013rowclone}.  Each
subarray contains {many} rows of DRAM cells, where each column of
DRAM cells is connected together across the multiple rows using
\emph{bitlines}.

RowClone consists of two mechanisms that take advantage of the
existing DRAM structure.  The first mechanism, Fast Parallel Mode,
copies {the data of a row inside a subarray to another row}
inside the same DRAM subarray by issuing back-to-back activate (i.e.,
row open) commands to the source and the destination row.  The second
mechanism, Pipelined Serial Mode, {can transfer an arbitrary
  number of bytes} between two banks using the shared internal bus
among banks in a DRAM chip.

RowClone significantly reduces the raw latency and energy consumption
of bulk data copy and initialization, leading to $11.6\times$ latency
reduction and $74.4\times$ energy reduction for a {4kB bulk page
  copy (using the Fast Parallel Mode)}, at very low cost (only 0.01\%
DRAM chip area overhead)~\cite{seshadri2013rowclone}. This reduction
directly translates to improvement in performance and energy
efficiency of systems running copy or initialization-intensive
workloads.  {Our MICRO 2013 paper~\cite{seshadri2013rowclone}
  shows that the performance of six copy/{initialization}-{intensive}
  benchmarks (including {the} fork system call,
  Memcached~\cite{memcached} and a MySQL~\cite{mysql} database)
  improves between 4\% and 66\%. For the same six benchmarks, RowClone
  reduces the energy consumption between 15\% and 69\%.}

\subsubsection{Ambit}
\label{sec:ambit}

In addition to bulk data movement, many applications trigger
\emph{bulk bitwise operations}, i.e., bitwise operations on large bit
vectors~\cite{btt-knuth,hacker-delight}.  Examples of such
{applications} include bitmap
indices~\cite{bmide,bmidc,fastbit,bicompression} and bitwise scan
acceleration~\cite{bitweaving} for databases, accelerated document
filtering for web search~\cite{bitfunnel}, DNA sequence
alignment~\cite{bitwise-alignment,xin.shd.bioinformatics15,alser.bioinformatics17},
encryption algorithms~\cite{xor1,xor2,enc1}, graph
processing~\cite{li.dac16}, and
networking~\cite{hacker-delight}. Accelerating bulk bitwise operations
can {thus} significantly boost the performance and energy
efficiency of a wide range applications.

In order to avoid data movement bottlenecks when the system performs
these {bulk} bitwise operations, we have recently proposed a new
\textbf{A}ccelerator-in-\textbf{M}emory for bulk \textbf{Bit}wise
operations (Ambit)~\cite{Seshadri:2015:ANDOR, seshadri.arxiv16,
  seshadri.micro17}.  Unlike prior approaches, Ambit uses the analog
operation of existing DRAM technology to perform bulk bitwise
operations.  Ambit consists of two components.  The first component,
Ambit--AND--OR, implements a new operation called \emph{triple-row
  activation}, where the memory controller simultaneously activates
three rows.  Triple-row activation performs a bitwise majority
function across the cells in the three rows, {due to the charge
  sharing principles that govern the operation of the DRAM array}.  By
controlling the initial value of one of the three rows, we can use
triple-row activation to perform a bitwise AND or OR of the other two
rows.  The second component, Ambit--NOT, takes advantage of the two
inverters that are connected to each sense amplifier in a DRAM
subarray. {Ambit--NOT exploits the fact that, at the end of the
  sense amplification process, the voltage level of one of the
  inverters represents the negated logical value of the cell.  The
  Ambit design adds a special row to the DRAM array, which is used to
  capture the negated value that is present in the sense amplifiers.
  One possible implementation of the special
  row~\cite{seshadri.micro17} is a row of \emph{dual-contact cells} (a
  2-transistor 1-capacitor cell~\cite{2t-1c-1,migration-cell}) that
  connects to both inverters inside the sense amplifier.}  With the
ability to perform AND, OR, and NOT operations, Ambit {is
  functionally complete: It} can reliably perform \emph{any} bulk
bitwise operation completely using DRAM technology, even in the
presence of significant process variation
{(see \cite{seshadri.micro17} for details)}.

Averaged across seven commonly-used bitwise operations, Ambit with 8
DRAM banks improves bulk bitwise operation throughput by 44$\times$
compared to an Intel Skylake processor~\cite{intel-skylake}, and
32$\times$ compared to the NVIDIA GTX 745 GPU~\cite{gtx745}. Compared
to the DDR3 {standard}, Ambit reduces energy consumption of these
operations by 35$\times$ on average. Compared to HMC
2.0~\cite{hmc.spec.2.0}, Ambit improves bulk bitwise operation
throughput by 2.4$\times$. When integrated directly into the HMC 2.0
device, Ambit improves throughput by 9.7$\times$ compared to
processing in the logic layer of HMC 2.0.

{A number of Ambit-like bitwise operation substrates have been
  proposed in recent years, making use of emerging resistive memory
  technologies, e.g., {phase-change memory}
  (PCM)~\cite{lee-isca2009,lee.ieeemicro10,lee.cacm10,zhou.isca09,qureshi.isca09,yoon-taco2014},
  SRAM, or specialized computational DRAM. These substrates can
  perform bulk bitwise operations in a special DRAM array augmented
  with computational circuitry~\cite{li.micro17} and in 
  {PCM}~\cite{li.dac16}. Similar substrates can perform simple
  arithmetic operations in SRAM~\cite{aga.hpca17, kang.icassp14} and
  arithmetic and logical operations in
  memristors~\cite{kvatinsky.tcasii14, kvatinsky.iccd11,
    kvatinsky.tvlsi14, shafiee.isca16, levy.microelec14}. We believe
  it is extremely important to continue exploring such low-cost
  Ambit-like {substrates,} as well as more sophisticated computational substrates{,}
  for all types of memory technologies, old and new. Resistive memory
  technologies are fundamentally non-volatile and amenable to in-place
  updates, and as such, can lead to even less data movement compared
  to DRAM, which fundamentally requires some data movement to access
  the data. {Thus}, we believe it is very promising to examine the
  design of emerging resistive memory chips that can incorporate
  Ambit-like bitwise {operations} and other types of suitable computation
  capability.}

\subsubsection{Gather-Scatter DRAM}
\label{sec:gs-dram}

Many applications access data structures with different access
patterns {at different points in time}. Depending on the layout
of the data structures in the physical memory, some access patterns
require non-unit strides. As current memory systems are optimized to
access {sequential} cache lines, non-unit strided accesses
exhibit low spatial locality, {leading to memory bandwidth
  waste} and cache space waste.

Gather-Scatter DRAM (GS-DRAM)~\cite{GS-DRAM} is a low-cost substrate
that addresses this problem. It {performs} in-DRAM data structure
reorganization by accessing multiple values that belong to a strided
access pattern using a single read/write command {in the memory
  controller}. GS-DRAM uses two key new mechanisms.  First, GS-DRAM
remaps the data of each cache line to different chips such that
multiple values of a strided access pattern are mapped to different
chips.  {This enables the possibility of gathering different
  parts of the strided access pattern concurrently from different
  chips.}  Second, instead of sending separate requests to each chip,
the GS-DRAM memory controller communicates a pattern ID to the memory
module. With the pattern ID, each chip computes the address to be
accessed independently. This way, the returned cache line contains
different values of the strided pattern gathered from different chips.

GS-DRAM achieves near-ideal memory bandwidth and cache utilization in
real-world workloads, such as in-memory databases and matrix
multiplication.  {For in-memory databases, GS-DRAM outperforms a
  transactional workload with column store layout by $3\times$ and an
  analytics workload with row store layout by $2\times$, thereby
  getting the best performance {for} both transactional and analytical
  queries on databases (which in general benefit from different types
  of data layouts).  For matrix multiplication, GS-DRAM is 10\% faster
  than the best-performing tiled implementation of {the matrix
  multiplication algorithm}.}

\subsection{Approach II: PIM using 3D-Stacked Memory}
\label{sec:3dstacked}

Several works propose to place some form of processing logic
(typically accelerators, simple cores, or reconfigurable logic) inside
the logic layer of 3D-stacked memory~\cite{lee.taco16}.  This
\emph{PIM processing logic}, which we also refer to as \emph{PIM
  cores} or \emph{PIM engines}, interchangeably, can execute portions
of applications (from individual instructions to functions) or entire
{threads and applications,} depending on the design of the
architecture. Such PIM engines have high-bandwidth and low-latency
access to the memory stacks that {are on} top of them, since the logic
layer and the memory layers are connected via high-bandwidth vertical
connections~\cite{lee.taco16}, e.g., through-silicon vias. In this
section, we discuss how systems can make use of relatively simple PIM
engines within the logic layer to avoid data movement and {thus
  obtain} significant performance and energy {improvements} on a
wide variety of application domains.

\subsubsection{Tesseract: Graph Processing}
\label{sec:tesseract}

A {popular modern} application is large-scale graph
processing~\cite{salihoglu.ssdbm13, tian.vldb13, low.vldb12,
  hong.asplos12, malewicz.sigmod10, harshvardhan.pact14,
  gonzalez.osdi12, ligra, Seraph, graphlab, nai2017graphpim}.
{Graph processing has} broad applicability and use in many
domains, from social {networks} to machine learning, from
{data analytics} to bioinformatics.  Graph analysis workloads
are known to put {significant} pressure on memory bandwidth due
to (1) large amounts of random memory accesses across large memory
regions (leading to very limited cache efficiency and very large
amounts of unnecessary data transfer on the memory bus) and (2) very
small amounts of computation per each data item {fetched from
  memory} (leading to very limited ability to hide long memory
latencies {and exacerbating the energy bottleneck by exercising
  the huge energy disparity between memory access and computation}).
These two characteristics make it very challenging to scale up such
workloads despite their inherent parallelism, especially with
conventional architectures based on large on-chip caches and
{relatively} scarce off-chip memory bandwidth {for random
  access}.

We can exploit the high bandwidth as well as {the} potential computation
capability available within the logic layer of 3D-stacked memory to
overcome the limitations of conventional architectures for graph
processing.  To this end, we design a programmable PIM accelerator for
large-scale graph processing, called
Tesseract~\cite{ahn.tesseract.isca15}.  Tesseract consists of (1) a
new hardware architecture that {effectively} utilizes the
available memory bandwidth in 3D-stacked memory by placing simple
in-order processing cores in the logic layer and enabling each core
{to manipulate} data only on the memory partition it is assigned to
control, (2) an efficient method of communication between different
in-order cores within a 3D-stacked memory to enable each core to
request computation on data elements that reside in the memory
partition controlled by another core, and (3) a {message-passing
  based} programming interface, {similar to how modern
  distributed systems are programmed}, which enables remote function
calls on data that resides in each memory partition. The Tesseract
design moves functions to data rather than moving data elements across
different memory partitions and cores. It also includes two hardware
prefetchers specialized for memory access patterns of graph
processing, which operate based on the hints provided by our
programming model.  Our comprehensive evaluations using five
state-of-the-art graph processing workloads with large real-world
graphs show that the proposed Tesseract PIM architecture improves
average system performance by {$13.8\times$} and achieves 87\%
average energy reduction over conventional systems.

\subsubsection{Consumer Workloads}
\label{sec:google}

A very popular domain of computing today consists of consumer devices,
which include smartphones, tablets, web-based computers such as
Chromebooks, and wearable devices.  In consumer devices, energy
efficiency is a first-class concern due to the limited battery
capacity and {the stringent} thermal power budget.  We find that
\emph{data movement} is a major contributor to the total system energy
     {and execution time} in modern consumer devices.  Across all of
     the popular modern applications we study (described in the next
     paragraph), we find that a massive 62.7\% of the total system
     energy, on average, is spent on data movement {across the
       memory hierarchy~\cite{boroumand.asplos18}.}

We comprehensively analyze the energy and performance impact of data
movement for several widely-used Google consumer
workloads~\cite{boroumand.asplos18}, which account for a significant
portion of the applications executed on consumer devices.  These
workloads include (1)~{the Chrome web browser}~\cite{chrome}, which is
a very popular browser used in mobile devices and laptops;
(2)~{TensorFlow Mobile}~\cite{mobile-tensorflow}, {Google's machine
  learning framework, which is used in services such as Google
  Translate, Google Now, and Google Photos; (3)~{the VP9 video
    playback engine}~\cite{vp9-specification}, and (4)~{the VP9
    video capture engine}~\cite{vp9-specification}, both of which are
  used in many video services such as YouTube and Google Hangouts.  We
  find that {offloading key functions to the logic layer} can
  greatly reduce data movement in all of these workloads.  However,
  there are challenges to introducing PIM in consumer devices, as
  consumer devices are extremely stringent in terms of the area and
  energy budget {they can accommodate for any new hardware
    enhancement}.  As a result, we need to identify what kind of
  in-memory logic can both (1)~\emph{maximize energy efficiency} and
  (2)~be implemented at \emph{minimum possible cost, in terms of both
    area overhead and complexity}.

We find that many of target functions for PIM in consumer workloads
are comprised of simple operations such as \emph{memcopy}, \emph{memset},
basic arithmetic and bitwise operations, {and simple data
  shuffling and reorganization routines}.  Therefore, we can
{relatively easily} implement these PIM target functions in
{the logic layer of 3D-stacked} memory using either (1)~a small
low-power general-purpose embedded core or (2)~a group of small
fixed-function accelerators.  Our analysis shows that the area of a
PIM core and a PIM accelerator take up no more than 9.4\% and 35.4\%,
respectively, of the area available for PIM logic in an
HMC-like~\cite{hmcspec2} 3D-stacked memory architecture.  Both the PIM
core and PIM accelerator eliminate a large amount of data movement,
and {thereby} significantly reduce total system energy (by an
average of 55.4\% across {all} the workloads) and execution time
(by an average of 54.2\%).

\subsubsection{GPU Applications}
\label{sec:gpu}

In the last decade, Graphics Processing Units (GPUs) have become the
accelerator of choice for {a wide variety of} data-parallel
applications. They deploy thousands of in-order, {SIMT (Single
  Instruction Multiple Thread)} cores that run lightweight
threads. Their multithreaded architecture is devised to hide the long
latency of memory accesses by interleaving threads that execute
arithmetic and logic operations. Despite that, many GPU applications
are {still} very
memory-bound~\cite{veynu.2011,jog.asplos2013,jog.isca2013,medic,nandita.2015,nandita.2016,jog.2016,ausavarungnirun.asplos18,ausavarungnirun.micro17,rachata-thesis},
because the limited off-chip pin bandwidth cannot supply enough data
to the running threads.

3D-stacked memory architectures present a promising opportunity to
alleviate the memory bottleneck in GPU systems. GPU cores placed
{in the logic layer of a 3D-stacked memory} can be directly
connected to the DRAM layers with high bandwidth (and low latency)
connections.  In order to leverage the potential performance benefits
of such systems, it is necessary to enable computation offloading and
data mapping to multiple {such compute-capable} 3D-stacked
memories, {such} that GPU applications can benefit from
processing-in-memory capabilities in the logic layers {of such
  memories}.

TOM (Transparent Offloading and Mapping)~\cite{hsieh.isca16} proposes
two mechanisms to address computation offloading and data mapping
{in such a system} in a programmer-transparent manner.  First,
it introduces {new} compiler analysis {techniques} to
identify code sections in GPU kernels that can benefit from PIM
offloading. The compiler estimates the potential memory bandwidth
savings for each code block. To this end, {the compiler} compares
the bandwidth consumption of the code block, when executed on the
regular GPU cores, to the bandwidth cost of transmitting/receiving
input/output registers, when offloading to the GPU cores in the logic
layers. {At runtime, a final} offloading decision is made based
on system conditions, such as contention for processing resources in
the logic layer.  Second, a software/hardware {cooperative}
mechanism predicts the memory pages that will be accessed by offloaded
code, and places {such pages} in the same 3D-stacked memory cube
where the code will be executed.  The goal is to make PIM effective by
ensuring that the data needed by the PIM cores is in the same memory
stack.  {Both mechanisms are completely transparent to the
  programmer, who only needs to write regular GPU code without any
  explicit PIM instructions or any other modification to the code.}
TOM improves the average performance {of a variety of GPGPU
  workloads} by 30\% and reduces the average energy consumption by
11\% with respect to a baseline GPU system without PIM offloading
capabilities.

A related work~\cite{pattnaik.pact16} identifies GPU kernels that are
suitable for PIM offloading by using a regression-based affinity
prediction model. A concurrent kernel management mechanism uses the
affinity prediction model and determines which kernels {should}
be scheduled concurrently {to maximize performance}. This way,
{the proposed} mechanism enables the simultaneous exploitation
of the regular GPU cores and the in-memory GPU cores. {This
  scheduling technique improves} performance and energy efficiency by
an average of 42\% and 27\%, respectively.

\subsubsection{PEI: PIM-Enabled Instructions}
\label{sec:pei}

PIM-Enabled Instructions (PEI)~\cite{ahn.pei.isca15} aims to provide
the minimal processing-in-memory support to take advantage of PIM
using 3D-stacked memory, in a way that can achieve significant
performance and energy benefits without changing the computing system
significantly.  To this {end}, PEI proposes a collection of
simple instructions, which introduce negligible changes to the
computing system and no changes to the programming model or the
virtual memory system, in a system with 3D-stacked memory. These
instructions, inserted by the compiler/programmer to code written in a
regular program, are operations that can be executed either in a
traditional host CPU (that fetches and decodes them) or the PIM engine
in 3D-stacked memory.

PIM-Enabled Instructions are based on two key ideas. First, a PEI is a
cache-coherent, virtually-addressed host processor instruction that
operates on only a single cache block. It requires no changes to the
sequential execution and programming model, no changes to virtual
memory, minimal changes to cache coherence, and no need for special
data mapping to take advantage of PIM (because each PEI is restricted
to a single memory module due to the single cache block restriction it
has). Second, a Locality-Aware Execution runtime mechanism decides
dynamically where to execute a PEI (i.e., {either} the host
processor or the PIM logic) based on simple locality characteristics
and simple hardware predictors. {This runtime mechanism executes
  the PEI at the location that maximizes performance.}  In summary,
PIM-Enabled Instructions provide the illusion that PIM operations are
executed as if they were host instructions.

Examples of PEIs are integer increment, integer minimum,
floating-point addition, hash table probing, histogram bin index,
Euclidean distance, and dot product~\cite{ahn.pei.isca15}.
Data-intensive workloads such as graph processing, in-memory data
analytics, machine learning, and data mining can {significantly}
benefit from these PEIs.  Across 10 {key data-intensive}
workloads, we observe {that the use of PEIs in these workloads,
  in combination with the Locality-Aware Execution runtime mechanism,
  leads to an} average performance improvement of 47\% and an average
energy reduction of 25\% over a baseline {CPU}.

\section{Enabling the Adoption of PIM}
\label{sec:adoption}

Pushing some or all of the computation for a program from the CPU to
{memory} introduces new challenges for system architects
{and programmers} to overcome.  These challenges must be
addressed in order for PIM to be adopted as a mainstream architecture
in a wide variety of systems and workloads, and in a seamless manner
that does not place heavy burden on the vast majority of programmers.
In this section, we discuss several of these system-level {and
  programming-level} challenges, and highlight a number of our works
that have addressed these challenges for a wide range of PIM
architectures.

\subsection{Programming Model and Code Generation}
\label{sec:mapping}

Two open research questions to enable the adoption of PIM are 1) what
should the programming {models} be, and 2) how can compilers and
libraries alleviate the programming burden.

While PIM-Enabled Instructions~\cite{ahn.pei.isca15} work well for
offloading small amounts of computation to memory, they can
potentially {introduce overheads while taking} advantage of PIM
for large tasks, due to the need to frequently exchange information
between {the PIM processing logic} and the CPU.  Hence, there is
a need for researchers to investigate how to integrate PIM
instructions with other compiler-based methods or library calls that
can support PIM integration, and how these approaches can ease the
burden on the programmer, by enabling seamless offloading of
instructions or function/library calls.

Such solutions can often be platform-dependent.  One of our recent
works~\cite{hsieh.isca16} examines compiler-based mechanisms to decide
what portions of code should be offloaded to PIM processing logic in a
GPU-based system in a manner that is transparent to the GPU
programmer.  Another recent work~\cite{pattnaik.pact16} examines
system-level techniques that decide which GPU application kernels are
suitable for PIM execution.

Determining {effective programming interfaces} and the necessary
compiler/library support to perform PIM remain open research
{and design questions, which are important for future works to
  tackle.}

\subsection{PIM Runtime: Scheduling and Data Mapping}
\label{sec:scheduling}

We identify four key runtime issues in PIM: (1)~what {code} to
execute near data, (2)~when to schedule execution on PIM (i.e., when
is it worth offloading computation to the PIM cores), (3)~how to map
data to multiple memory modules such that PIM execution is viable and
effective, and (4)~how to effectively share/partition PIM
mechanisms/accelerators at runtime across multiple threads/cores to
maximize performance and energy efficiency.  We have already proposed
several approaches to solve these four issues.

Our recent works in PIM processing identify suitable \emph{PIM
  offloading candidates} with different granularities. PIM-Enabled
Instructions~\cite{ahn.pei.isca15} {propose various} operations
that can benefit from execution near or inside memory, such as integer
increment, integer minimum, floating-point addition, hash table
probing, histogram bin index, Euclidean distance, and dot product.
In~\cite{boroumand.asplos18}, we find {simple} functions with
intensive data movement that are suitable for PIM in consumer
workloads (e.g., Chrome web browser, TensorFlow Mobile, video
playback, and video capture), {as described in
  Section~\ref{sec:google}}.  Bulk memory operations (copy,
{initialization}) and {bulk} bitwise operations are good candidates
for in-DRAM
processing{~\cite{seshadri2013rowclone,seshadri.micro17,Seshadri:2015:ANDOR, seshadri.thesis16}}.
GPU applications also contain several parts that are suitable
{for offloading to PIM
  engines}~\cite{hsieh.isca16,pattnaik.pact16}.

In several of our research works, we propose runtime mechanisms for
\emph{dynamic scheduling} of PIM offloading candidates, i.e.,
mechanisms that decide whether or not to actually offload code that is
marked to be potentially offloaded to PIM engines.
In~\cite{ahn.pei.isca15}, we develop a locality-aware scheduling
mechanism {for PIM-enabled instructions}.  For GPU-based
systems~\cite{hsieh.isca16,pattnaik.pact16}, we explore the
combination of compile-time and runtime mechanisms for identification
and dynamic scheduling of PIM offloading candidates.

The best \emph{mapping of data and code} that enables the maximal
benefits from PIM depends on the applications and the computing system
configuration. For instance, {in~\cite{hsieh.isca16}, we present}
a software/hardware mechanism to map data and code to several
3D-stacked memory cubes in regular GPU applications with relatively
regular memory access patterns. This work also deals with effectively
\emph{sharing PIM {engines} across multiple threads}, as GPU
code sections are offloaded from different GPU cores.  Developing
{new approaches} to data/code mapping and scheduling for a large
variety of applications and possible core and memory configurations is
still necessary.

In summary, there are still several key research questions that should
be investigated {in runtime systems for PIM, which perform
  scheduling and data/code mapping}:
\begin{itemize}
\item What are simple mechanisms to enable and disable PIM execution? How can PIM execution be throttled for highest performance gains? How should data locations and access patterns affect where/whether PIM execution should occur?
\item Which parts of {a given application's} code should be executed on PIM? What are simple mechanisms to identify when those parts of the application code can benefit from PIM?
\item What are scheduling mechanisms to share PIM {engines} between multiple requesting cores to maximize {benefits obtained from PIM}? 
\item What are simple mechanisms to {manage access to a memory that serves both CPU requests and PIM requests?}
\end{itemize}

\subsection{Memory Coherence}
\label{sec:coherence}

In a traditional multithreaded execution model that makes use of
shared memory, writes to memory must be coordinated between multiple
CPU cores, to ensure that threads do not operate on stale data values.
Since CPUs include per-core private caches, when one core writes data
to a memory address, cached copies of the data held within the caches
of other cores must be updated or invalidated, using a
{mechanism} known as \emph{cache coherence}.  Within a modern
chip multiprocessor, the per-core caches perform coherence actions
over a shared interconnect, with hardware coherence protocols.

Cache coherence is a major system challenge for enabling PIM
architectures as general-purpose execution engines, as {PIM
  processing logic} can modify the data {it processes}, and this
data may also be needed by CPU cores.  If {PIM processing logic
  is} coherent with the processor, the PIM programming model is
relatively simple, as it remains similar to conventional {shared
  memory} multithreaded programming{, which makes PIM architectures
  easier to adopt in general-purpose systems}.  Thus, allowing
{PIM processing logic} to maintain such a simple and traditional
shared memory programming model can facilitate the {widespread}
adoption of PIM.  However, employing traditional fine-grained cache
coherence {(e.g., a cache-block based MESI
  protocol~\cite{mesi1984})} for PIM forces a large number of
coherence messages to traverse {the narrow processor-memory
  bus}, potentially undoing the benefits of high-bandwidth {and
  low-latency} PIM {execution}.  {Unfortunately,} solutions
for coherence proposed by prior PIM
works~\cite{ahn.pei.isca15,hsieh.isca16} either place some
restrictions on the programming model (by eliminating coherence and
requiring message passing based programming) or limit the performance
{and energy} gains achievable by a PIM architecture.

We have developed a new coherence protocol,
LazyPIM~\cite{boroumand2016pim,boroumand.arxiv17}, that maintains
cache coherence between PIM processing logic and CPU cores
\emph{without} sending coherence requests for every memory access.
Instead, LazyPIM efficiently provides coherence by having PIM
processing logic \emph{speculatively} acquire coherence permissions,
and then later sends compressed \emph{batched} coherence lookups to
the CPU to determine whether or not its speculative permission
acquisition violated the {coherence semantics.}  As a result
{of this "lazy" checking of coherence violations}, LazyPIM
approaches near-ideal coherence behavior: {the performance and
  energy consumption of a PIM architecture with LazyPIM are,
  respectively, within 5.5\% and 4.4\% the performance and energy
  consumption of} a system where coherence is performed at zero
latency and energy cost.

{Despite the leap that
  LazyPIM~\cite{boroumand2016pim,boroumand.arxiv17} represents for
  memory coherence in computing systems with PIM support, we believe
  that it is still necessary to explore other solutions for memory
  coherence that can efficiently deal with all types of workloads and
  PIM offloading granularities.}

\subsection{Virtual Memory Support}
\label{sec:virtualmemory}

When an application needs to access its data inside the main memory,
the CPU core must first perform an \emph{address translation}, which
converts the data's virtual address into a \emph{physical} address
within main memory.  If the translation {metadata} is not
available in the CPU's translation lookaside buffer (TLB), the CPU
must invoke the page table walker in order to perform a long-latency
page table walk that involves multiple \emph{sequential} reads to the
main memory and lowers the application's performance. In modern
systems, the virtual memory system also provides access protection
mechanisms.

A naive solution to reducing the overhead of page walks is to utilize
PIM engines to perform page table walks. This can be done by
duplicating the content of the TLB and {moving} the page walker
{to} the PIM processing logic in main memory.  Unfortunately,
this is either difficult or expensive for three reasons. First,
coherence {has} to be maintained between the CPU's TLBs and the
memory-side TLBs. This introduces extra complexity and off-chip
requests.  Second, duplicating the TLBs increases the storage and
complexity overheads {on the memory side, which should be
  carefully contained}.  Third, if main memory is shared across
{CPUs with} different types of architectures, page table
structures and the implementation of address translations can be
different across {the different} architectures. Ensuring
compatibility between the in-memory TLB/page walker and all possible
types of {virtual memory} architecture designs can be
complicated and often not even practically feasible.

To address these concerns and reduce the overhead of virtual memory,
we explore a tractable solution for PIM address translation as part of
our in-memory pointer chasing accelerator, IMPICA~\cite{impica}.
IMPICA exploits the high bandwidth available within 3D-stacked memory
to traverse a chain of virtual memory pointers within DRAM,
\emph{without} having to look up virtual-to-physical address
translations in the CPU translation lookaside buffer (TLB) and without
using the page walkers within the CPU.  {IMPICA's key ideas are
  1) to use a region-based page table, which is optimized for PIM
  acceleration, and 2) to decouple address calculation and memory
  access with two specialized engines.  IMPICA improves the
  performance of pointer chasing operations in three commonly-used
  linked data structures (linked lists, hash tables, and B-trees) by
  92\%, 29\%, and 18\%, respectively. On a real database application,
  DBx1000, IMPICA improves transaction throughput and response time by
  16\% and 13\%, respectively. IMPICA also reduces overall system
  energy consumption (by 41\%, 23\%, and 10\% for the three
  commonly-used data structures, and by 6\% for DBx1000).}

Beyond pointer chasing operations that are tackled by
IMPICA~\cite{impica}, providing efficient mechanisms for PIM-based
virtual-to-physical address translation (as well as access protection)
remains a challenge for the generality of applications,
{especially those that access large amounts of virtual
  memory~\cite{mask,mosaic,mosaic-osr}.  We believe it is important to
  explore new ideas to address this PIM challenge in a {scalable and}
  efficient manner.}

\subsection{Data Structures for PIM}
\label{sec:datastructures}

Current systems with many cores run applications with concurrent data
structures to achieve high performance and scalability, with
significant benefits over sequential data structures. Such concurrent
data structures {are often used in} 
heavily-optimized server systems today, where high performance is
critical. To enable the adoption of PIM in such many-core systems, it
is necessary to develop concurrent data structures that are
specifically tailored to take advantage of PIM.

\emph{Pointer chasing data structures} and \emph{contended data
  structures} require careful analysis and design to leverage the high
bandwidth and low latency of 3D-stacked memories~\cite{liu-spaa17}.
First, pointer chasing data structures, such as linked-lists and
skip-lists, have a high degree of inherent parallelism and low
contention, but a naive implementation in PIM cores is burdened by
hard-to-predict memory access patterns. By combining and partitioning
the data across 3D-stacked memory vaults, it is possible to fully
exploit the inherent parallelism of these data structures.  Second,
contended data structures, such as FIFO queues, are a good fit for CPU
caches because they expose high locality. However, they suffer from
high contention when many threads access them concurrently.
{Their performance on traditional CPU systems can be improved
  using a new PIM-based FIFO queue~\cite{liu-spaa17}.  The proposed
  PIM-based FIFO queue uses a PIM core to perform enqueue and dequeue
  operations requested by CPU cores. The PIM core can pipeline
  requests from different CPU cores for improved performance.}

{As {recent work~\cite{liu-spaa17}} shows, PIM-managed concurrent data
  structures can outperform state-of-the-art concurrent data
  structures that {are designed} for and executed on multiple cores. We
  believe and hope that future work will enable other types of data
  structures (e.g., hash tables, search trees, priority queues) to
  benefit from PIM-managed designs.}

\subsection{Benchmarks and Simulation Infrastructures}
\label{sec:simulation}

To ease the adoption of PIM, it is critical that we {accurately}
assess the benefits and shortcomings of PIM. Accurate assessment of
PIM requires (1)~a {preferably large} set of real-world
memory-intensive applications that have the potential to benefit
significantly when executed near memory, (2)~a rigorous methodology to
(automatically) identify PIM offloading candidates, and
(3)~simulation/evaluation infrastructures that allow architects and
system designers to {accurately} analyze the benefits and
overheads of adding PIM processing logic to memory and executing code
on this processing logic.

In order to explore what processing logic should be introduced near
memory, and to know what properties are ideal for PIM kernels, we
believe it is important to begin by developing a \emph{real-world
  benchmark suite} of a wide variety of applications that can
potentially benefit from PIM. While many data-intensive applications,
such as pointer chasing and bulk memory copy, can potentially benefit
from PIM, it is crucial to examine important candidate applications
for PIM execution, and for researchers to agree on a common set of
these candidate applications to focus the efforts of the community
{as well as to enable reproducibility of results, which is
  important to assess the relative benefits of different ideas
  developed by different researchers}. We believe that these
applications should come from a number of popular and emerging
domains. Examples of potential domains include data-parallel
applications, neural networks, machine learning, graph processing,
data analytics, search/filtering, mobile workloads, bioinformatics,
Hadoop/Spark programs, security/cryptography, and in-memory data
stores. Many of these applications have large data sets and can
benefit from high memory bandwidth and low memory latency benefits
provided by computation near memory. In our prior work, we have
started identifying several applications that can benefit from PIM in
graph processing
frameworks~\cite{ahn.pei.isca15,ahn.tesseract.isca15}, pointer
chasing~\cite{cont-runahead,impica}, databases~\cite{boroumand2016pim,
  boroumand.arxiv17, impica, GS-DRAM}, consumer
workloads~\cite{boroumand.asplos18}, machine
learning~\cite{boroumand.asplos18}, and GPGPU
workloads~\cite{hsieh.isca16,pattnaik.pact16}.  {However, there
  is significant room for methodical development of a large-scale PIM
  benchmark suite, which we hope future work provides.}

A systematic \emph{methodology} for (automatically) identifying
potential PIM kernels (i.e., code portions that can benefit from PIM)
within an application can, {among many other benefits, 1) ease
  the burden of programming PIM architectures by aiding the programmer
  to identify what should be offloaded, 2) ease the burden of and
  improve the reproducibility of PIM research, 3) drive the design and
  implementation of PIM functional units that many types of
  applications can leverage, 4) inspire the development of tools that
  programmers and compilers can use to automate the process of
  offloading portions of existing applications to PIM processing
  logic, and 5) lead the community towards convergence on PIM designs
  and offloading candidates.}

{We also} need \emph{simulation infrastructures} to accurately
model the performance and energy of PIM hardware structures, available
memory bandwidth, and communication overheads when we execute code
near or inside memory. Highly-flexible and commonly-used memory
simulators (e.g., Ramulator~\cite{ramulator, ramulator.github},
SoftMC~\cite{hassan2017softmc, softmc.github}) can be combined with
full-system simulators (e.g., gem5~\cite{GEM5}, zsim~\cite{zsim},
gem5-gpu~\cite{gem5-gpu}, GPGPUSim~\cite{gpgpusim}) to provide a
robust environment that can evaluate how various PIM architectures
affect the \emph{entire compute stack}, and can allow designers to
identify memory, workload, and system characteristics that affect the
efficiency of PIM execution. We believe it is critical to support the
open source development such {simulation and} emulation infrastructures
for assessing the benefits of a wide variety of PIM designs.

\section{Conclusion and Future Outlook}
\label{sec:conclusion}

Data movement is a major performance and energy bottleneck plaguing
modern computing systems. A large fraction of system energy is spent
on moving data across the memory hierarchy into the processors (and
accelerators), the only place where computation is performed in a
modern system. Fundamentally, the large amounts of data movement {are}
caused by the processor-centric design of modern computing systems:
processing of data is performed only in the processors (and
accelerators), which are far away from the data, and as a result, data
moves a lot in the system, to facilitate computation on it.

In this work, we argue for a paradigm shift in the design of computing
systems toward a data-centric design that enables computation
capability in places where data resides and thus performs computation
with minimal data movement.  {\emph{Processing-in-memory} (PIM) is
  a fundamentally data-centric design approach for computing systems
  that enables the ability to perform operations in or near memory.
  Recent advances in modern memory architectures have enabled us to
  extensively explore two novel approaches to designing PIM
  architectures.  First, with minimal changes to memory chips, we show
  that we can perform a number of important and widely-used operations
  (e.g., memory copy, data initialization, bulk bitwise operations,
  data reorganization) within DRAM.  Second, we demonstrate how
  embedded computation capability in the logic layer of 3D-stacked
  memory can be used in a variety of ways to provide significant
  performance improvements and energy savings, across a large range of
  application domains and computing platforms.}

{Despite the extensive design space that we have studied so far, a
  number of key challenges remain to enable the widespread adoption of
  PIM in future computing systems~\cite{ghose.bookchapter19,
    ghose.bookchapter19.arxiv}.}  {Important challenges include
  developing easy-to-use programming models for PIM (e.g., PIM
  application interfaces, compilers and libraries designed to abstract
  away PIM architecture details from programmers), and extensive
  runtime support for PIM (e.g., scheduling PIM operations, sharing
  PIM logic among CPU threads, cache coherence, virtual memory
  support).}  We {hope} that providing the community with (1)~a
large set of memory-intensive benchmarks that can potentially benefit
from PIM, (2)~a rigorous methodology to identify PIM-suitable parts
within an application, and (3)~accurate simulation infrastructures for
estimating the benefits and overheads of PIM will empower researchers
to address {remaining} challenges for the adoption of PIM.
  
{We firmly believe that it is time to design principled system
  architectures to solve the data movement problem of modern computing
  systems, which is caused by the rigid dichotomy and imbalance
  between the computing unit (CPUs and accelerators) and the
  memory/storage unit. Fundamentally solving the data movement problem
  requires a paradigm shift to a more data-centric computing system
  design, where computation happens in or near memory/storage, with
  minimal movement of data.  Such a paradigm shift can greatly push
  the boundaries of future computing systems, leading to orders of
  magnitude improvements in energy and performance (as we demonstrated
  with some examples in this work), potentially enabling new
  applications and computing {platforms.}}

\section*{{Acknowledgments}}

This work is based on a keynote talk delivered by Onur Mutlu at the
3rd Mobile System Technologies (MST) Workshop in Milan, Italy on 27
October 2017{~\cite{mutlu.msttalk17}}. The mentioned keynote talk
is similar to a series of talks given by Onur Mutlu in a wide variety
of venues since 2015 until now. This talk has evolved significantly
over time with the accumulationof new works and feedback received from
many audiences. A recent version of the talk was delivered as a
distinguished lecture at George Washington
University{~\cite{mutlu.gwutalk19}}.

This article and the associated talks are based on research done over
the course of the past seven years in the SAFARI Research Group on the
topic of {processing-in-memory} (PIM). We thank all of the members
of the SAFARI Research Group, and our collaborators at Carnegie
Mellon, ETH Zürich, and other universities, who have contributed to
the various works we describe in this paper. Thanks also goes to our
research group’s industrial sponsors over the past ten years,
especially Alibaba, Google, Huawei, Intel, Microsoft, NVIDIA, Samsung,
Seagate, and VMware. This work was also partially supported by the
Intel Science and Technology Center for Cloud Computing, the
Semiconductor Research Corporation, the Data Storage Systems Center at
Carnegie Mellon University, various NSF grants, and various awards,
including the NSF CAREER Award, the Intel Faculty Honor Program Award,
and a number of Google Faculty Research Awards to Onur Mutlu.

{
\bibliographystyle{elsarticle-num}
\bibliography{refs}
}
\end{document}